\documentclass[a4paper,
               bibtex,     
               keeplastbox,   
               ]{jacow}
\usepackage{pdfpages,multirow,ragged2e} %
\makeatletter%
	\ifboolexpr{bool{xetex}}
	 {\renewcommand{\Gin@extensions}{.pdf,%
	                    .png,.jpg,.bmp,.pict,.tif,.psd,.mac,.sga,.tga,.gif,%
	                    .eps,.ps,%
	                    }}{}
\makeatother

\ifboolexpr{bool{xetex} or bool{luatex}} 
 {}                                      
 {\usepackage[utf8]{inputenc}}           

\usepackage[UKenglish]{babel}

\ifboolexpr{bool{jacowbiblatex}}%
 {%
  \addbibresource{references.bib}
 }{}
\DeclareFontFamily{U}{BOONDOX-calo}{\skewchar\font=45 }
\DeclareFontShape{U}{BOONDOX-calo}{m}{n}{
  <-> s*[1.05] BOONDOX-r-calo}{}
\DeclareFontShape{U}{BOONDOX-calo}{b}{n}{
  <-> s*[1.05] BOONDOX-b-calo}{}
\DeclareMathAlphabet{\mathcalboondox}{U}{BOONDOX-calo}{m}{n}
\SetMathAlphabet{\mathcalboondox}{bold}{U}{BOONDOX-calo}{b}{n}
\DeclareMathAlphabet{\mathbcalboondox}{U}{BOONDOX-calo}{b}{n}
\newcommand{\calE}{\mathcalboondox{E}}

\newcommand{\mean}[1]{\langle #1 \rangle}

\let\OLDthebibliography\thebibliography
\renewcommand\thebibliography[1]{
  \OLDthebibliography{#1}
  \setlength{\parskip}{0pt}
  \setlength{\itemsep}{0pt plus 0.3ex}
}

\usepackage[scaled]{DejaVuSansMono}
\usepackage[T1]{fontenc}

\begin{document}

\title{\MakeLowercase{e}\textsuperscript{$+$}\MakeLowercase{e}\textsuperscript{$-$} Beam-beam Parameter Study for a T\MakeLowercase{e}V-scale PWFA Linear Collider}

\author{J. B. B. Chen\textsuperscript{1}, D. Schulte, CERN, Geneva, Switzerland \\
		E. Adli, University of Oslo, Oslo, Norway \\
		\textsuperscript{1}also at University of Oslo, Oslo, Norway}
	
\maketitle

\begin{abstract}
   We perform a beam-beam parameter study for a TeV-scale PWFA (particle-driven plasma wakefield acceleration) e\textsuperscript{$+$}e\textsuperscript{$-$} linear collider using GUINEA-PIG simulations. The study shows that the total luminosity follows the $1/\sqrt{\sigma_z}$-scaling predicted by beamstrahlung theory, where $\sigma_z$ is the rms beam length, which is advantageous for PWFA, as short beam lengths are preferred. We also derive a parameter set for a \SI{3}{\tera\electronvolt} PWFA linear collider with main beam parameters optimised for luminosity and luminosity spread introduced by beamstrahlung.
   
   Lastly, the study also compare the performance for scenarios with reduced positron beam charge at \SI{3}{\tera\electronvolt} and \SI{14}{\tera\electronvolt} with CLIC parameters.
\end{abstract}

\section{Introduction}

In the blow-out regime of PWFA (particle-driven plasma wakefield acceleration), a dense ultra-relativistic drive beam is used to excite a plasma wake, where plasma electrons are expelled from the region close to the propagation axis, leaving only positively charged ions behind to form a plasma ion bubble cavity. Inside the plasma ion cavity, accelerating gradients in the multi-\SI{}{\giga\volt/\meter} level \cite{Blumenfeld} can be used to accelerate a trailing main beam.

A previous parameter study on a \SI{1.5}{\tera\electronvolt} PWFA (particle-driven plasma wakefield acceleration) accelerator \cite{Chen2020modeling} derived a parameter set that can provide reasonable stability, energy spread and efficiency for electron acceleration. This study adopted the parameter set in \cite{Chen2020modeling}, assuming that positrons can be accelerated in a similar manner, and optimised the main beam parameters at the interaction point (IP) for a e\textsuperscript{$+$}e\textsuperscript{$-$} collider with respect to luminosity and luminosity spread introduced by beam-beam effects. The preference of short beams in PWFA can be exploited to reduce the beam sizes accordingly without increasing the level of beamstrahlung, while achieving a higher luminosity.

Furthermore, this study also examined asymmetric collision scenarios with reduced numbers of positrons at \SI{3}{\tera\electronvolt} and \SI{14}{\tera\electronvolt}.

\section{Beamstrahlung theory}
Colliding beams in a linear collider are focused to small transverse dimensions in order to reach high luminosity. This gives rise to intense electromagnetic fields that will bend the trajectories of particles in the opposite beam, and cause the particles to emit radiation in the form of beamstrahlung, and hence lose energy. A large fraction of particles will therefore collide with a less than nominal energy, and form a luminosity spectrum.


\subsection*{Beamstrahlung Parameter}
Beamstrahlung can be characterised by the critical energy defined at half power spectrum \cite{Schulte_thesis}
\begin{equation}
    \calE_\mathrm{c}=\hbar\omega_\mathrm{c}=\frac{3}{2}\frac{\hbar\gamma^3 c}{R},
\end{equation}
where $R$ is the bending radius of the particle trajectory.

It is however more convenient to use the dimensionless Lorentz invariant beamstrahlung parameter\index{Beamstrahlung parameter} defined as \cite{accHandbook, BeamBeam_Yokoya}
\begin{equation}
    \Upsilon = \frac{e\hbar}{m_\mathrm{e}^3c^3}(p_\mu F^{\mu\lambda} p^\nu F_{\lambda\nu})^{1/2},
\end{equation}
where $p^\mu$ is the four-momentum of the particle, and $F^{\mu\nu}$ is the electromagnetic field tensor of the beam field. The beamstrahlung parameter can also be written as
\begin{equation}
    \Upsilon = \frac{2}{3}\frac{\hbar\omega_\mathrm{c}}{\calE} = \gamma\frac{\mean{E+cB}}{B_\mathrm{c}},
\end{equation}
where $\calE$ is the energy of a particle before emitting radiation and $B_\mathrm{c} = m_\mathrm{e}^2c^2/(e\hbar) = \SI{4.4140}{\giga\tesla}$ is the Schwinger critical field. $\Upsilon$ can be interpreted as a measure for the strength of the electromagnetic fields in the rest frame of the electron in units of $B_\mathrm{c}$. Since fields above $B_\mathrm{c}$ are expected to cause nonlinear QED effects, $\Upsilon\ll 1$ is associated with the classical regime, while $\Upsilon\gg 1$ corresponds to the (deep) quantum regime.

$\Upsilon$ is not constant during collision. For Gaussian beams with $N$ particles, horizontal rms beam size $\sigma_x$, vertical rms beam size $\sigma_y$ and rms beam length $\sigma_z$, the average and maximum $\Upsilon$ can be approximated as
\begin{equation}
    \mean{\Upsilon} \approx \frac{5}{6}\frac{Nr_\mathrm{e}^2\gamma}{\alpha\sigma_z(\sigma_x+\sigma_y)} \quad \Upsilon_\mathrm{max}\approx\frac{12}{5}\mean{\Upsilon},
\end{equation}
where $r_\mathrm{e}$ is the classical electron radius and $\alpha$ is the fine structure constant.

\subsection*{Beamstrahlung and Luminosity}
In the quantum regime with $\Upsilon\gg 1$, the average number of emitted photons per electron during the collision for a Gaussian beam can be approximated as \cite{BeamBeam_Yokoya}
\begin{equation}
    n_\gamma\approx 2.54\frac{\alpha^2\sigma_z}{r_\mathrm{e}\gamma}\mean{\Upsilon}^{2/3} = 2.25\left( \frac{\alpha^2\sqrt{r_\mathrm{e}\sigma_z}N}{\sqrt{\gamma}(\sigma_x+\sigma_y)} \right)^{2/3}.
\label{eq:avgPhotonPerElectron}
\end{equation}
The total luminosity for a linear collider is given by
\begin{equation}
    \mathcal{L} = H_\mathrm{D}\frac{N^2}{4\pi\sigma_x\sigma_y}n_\mathrm{b}f_\mathrm{r} = H_\mathrm{D}\frac{N}{4\pi\sigma_x\sigma_y}\frac{P_\mathrm{b}}{\calE_\mathrm{b}},
\end{equation}
where $n_\mathrm{b}$ is the number of beams per pulse, $f_\mathrm{r}$ is the repetition rate of pulses, $P_\mathrm{b}=n_\mathrm{b}f_\mathrm{f}N\calE_\mathrm{b}$ is the beam power per beam, $\calE_\mathrm{b}$ is the beam energy and $H_\mathrm{D}$ is a correction factor usually in the range $1.5-2$ that takes into account the combined effect of the hourglass effect and disruption enhancement due to the
attractive force that the two colliding bunches exert on each other. Since $\mathcal{L}\propto 1/(\sigma_x\sigma_y)$ and $n_\gamma\propto 1/(\sigma_x+\sigma_y)^{2/3}$, choosing a flat beam with $\sigma_x\gg\sigma_y$ can limit $n_\gamma$ without sacrificing luminosity. This gives the following relation on $\sigma_x$ and $n_\gamma$: 
\begin{equation}
    \sigma_x = 3.38\frac{\alpha^2 N}{n_\gamma^{3/2}}\sqrt{\frac{r_\mathrm{e}\sigma_z}{\gamma}}.
\label{eq:sigmaX_quantum_beamstrahlung}
\end{equation}
Inserting this into the equation for the total luminosity, we obtain
\begin{equation}
    \mathcal{L} = \frac{0.30 H_\mathrm{D}}{4\pi\alpha^2} \sqrt{\frac{\gamma}{r_\mathrm{e}\sigma_z}} \frac{n_\gamma^{3/2}}{\sigma_y} \frac{\eta P_\mathrm{AC}}{\calE_\mathrm{b}},
\label{eq:luminosity_quantum_beamstrahlung}
\end{equation}
where $\eta$ is the total (wall-plug to beam) conversion efficiency, $P_\mathrm{AC}$ the wall-plug power for beam acceleration and $\calE_\mathrm{b}$ is the beam energy.

Equation \eqref{eq:avgPhotonPerElectron} shows that for $\Upsilon\gg 1$, a shorter beam can suppress beamstrahlung. This implies that $\sigma_x$ can be reduced accordingly for a flat beam, as described by equation \eqref{eq:sigmaX_quantum_beamstrahlung}, without increasing $n_\gamma$. Consequently, the luminosity can be increased for shorter beams, as outlined by equation \eqref{eq:luminosity_quantum_beamstrahlung}. This is particularly advantageous for PWFA, since short beams are preferred in PWFA due to the high plasma frequency. E.g. for a plasma with density $n_0=\SI{e16}{\centi\meter^{-3}}$, the plasma wavelength is $\lambda_\mathrm{p}=\SI{334}{\micro\meter}$. For comparison, $\lambda_\mathrm{RF}=\SI{2.51}{\centi\meter}$ in CLIC.



\section{Beam-beam parameter scan}
We performed beam-beam simulations using GUINEA-PIG \cite{Schulte_thesis}, where we optimised collisions of e\textsuperscript{$+$}e\textsuperscript{$-$} beams with respect to luminosity spread by performing parameter scans over $\beta_x$, $\beta_y$ and $\sigma_z$.

\subsection{Equal e\textsuperscript{$+$}e\textsuperscript{$-$} Beam Charges}

In this study, we assumed that the number of particles in both the e\textsuperscript{$+$} and the e\textsuperscript{$-$} beams are the same, and that $\beta_y$ can be made arbitrarily small regardless of technical constraints. Furthermore, we define the peak luminosity $\mathcal{L}_{0.01}$ as the part of the luminosity corresponding to centre of mass energy $\sqrt{s}>0.99\sqrt{s_0}$, where $\sqrt{s_0}$ is the nominal centre of mass collision energy. The acceptable level of luminosity spread is chosen to be $\mathcal{L}_{0.01}/\mathcal{L}\approx 1/3$, where $\mathcal{L}$ is the total luminosity.

\begin{figure}[!htb]
    \centering
    \includegraphics[width=0.9\columnwidth]{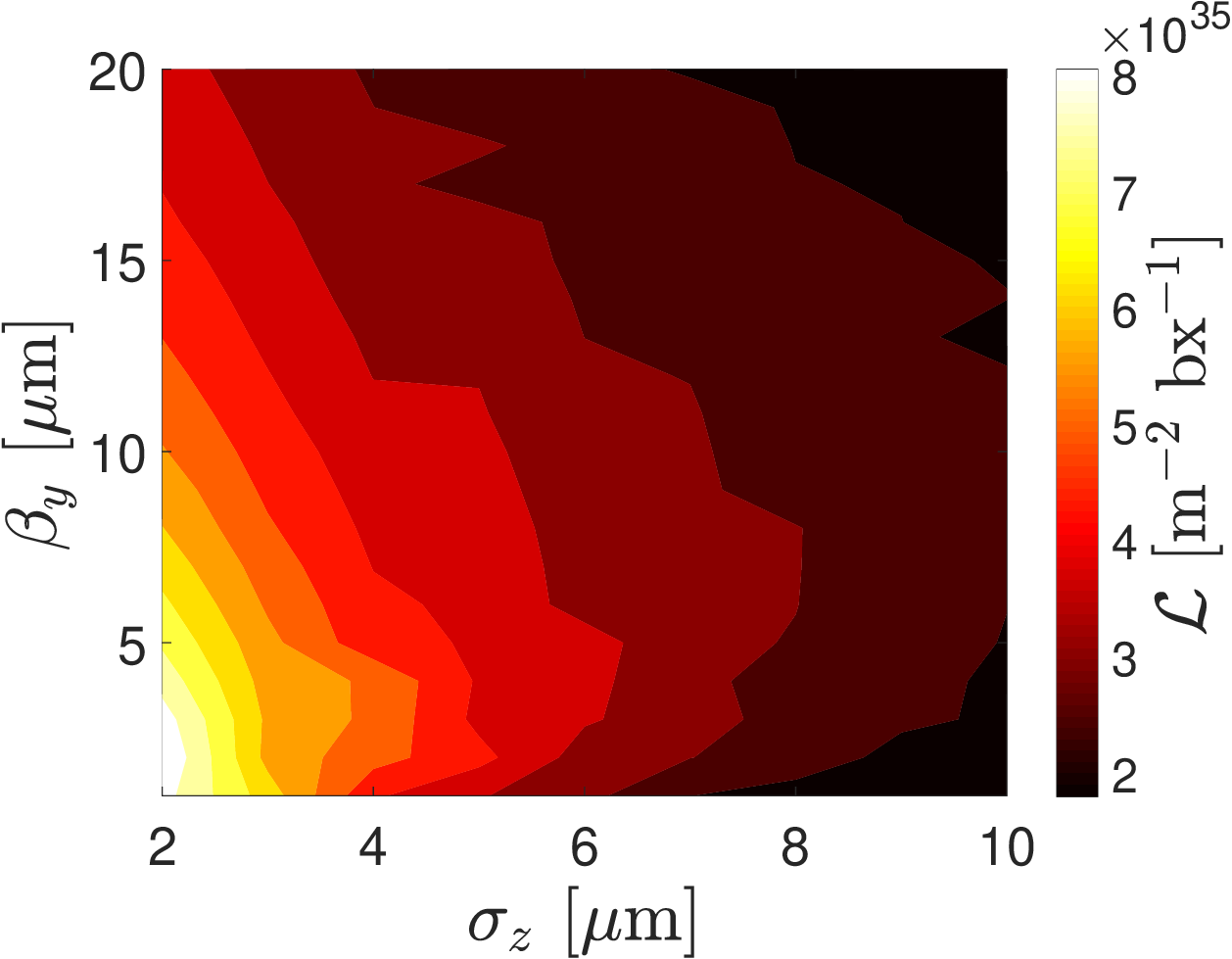}
    \caption{Contour plot of total luminosity $\mathcal{L}$ vs. beam length $\sigma_z$ and vertical beta function $\beta_y$, where the horizontal $\beta_x$ for each pair of $\sigma_z$ and $\beta_y$ has been chosen such that $\mathcal{L}_{0.01}/\mathcal{L}\approx 1/3$.}
    \label{fig:2020_N5e9_sigmaZ_betaY_lumi_contourPlot}
\end{figure}

\begin{figure}[!htb]
    \centering
    \includegraphics[width=0.9\columnwidth]{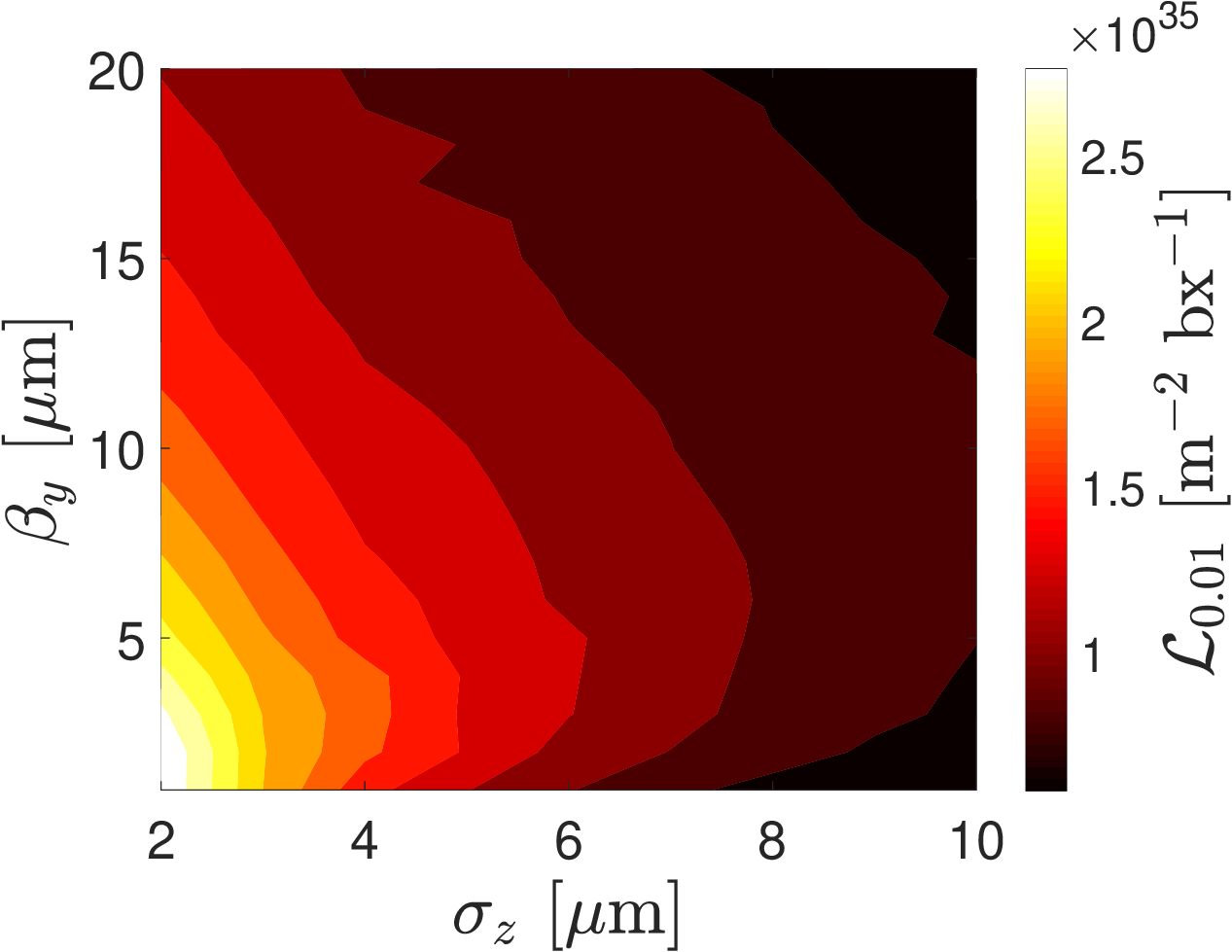}
    \caption{Contour plot of peak luminosity $\mathcal{L}_{0.01}$ vs. beam length $\sigma_z$ and vertical beta  function $\beta_y$, where the horizontal beta function $\beta_x$ for each pair of $\sigma_z$ and $\beta_y$ has been chosen such that $\mathcal{L}_{0.01}/\mathcal{L}\approx 1/3$.}
    \label{fig:2020_N5e9_sigmaZ_betaY_lumiHigh_contourPlot}
\end{figure}

Both beams have $N=\SI{5e9}{}$ particles and were collided at $\sqrt{s_0}=\SI{3}{\tera\electronvolt}$. For each pair of $\beta_y$ and $\sigma_z$, we kept only the results given by an optimal $\beta_x$ that corresponds to $\mathcal{L}_{0.01}/\mathcal{L}\approx 1/3$. The corresponding results for $\mathcal{L}$ and $\mathcal{L}_\mathrm{0.01}$ are shown in fig. \ref{fig:2020_N5e9_sigmaZ_betaY_lumi_contourPlot} and \ref{fig:2020_N5e9_sigmaZ_betaY_lumiHigh_contourPlot}, respectively. The unit bx\textsuperscript{$-1$} denotes ``per beam crossing''.

\begin{figure}[!htb]
    \centering
    \includegraphics[width=0.9\columnwidth]{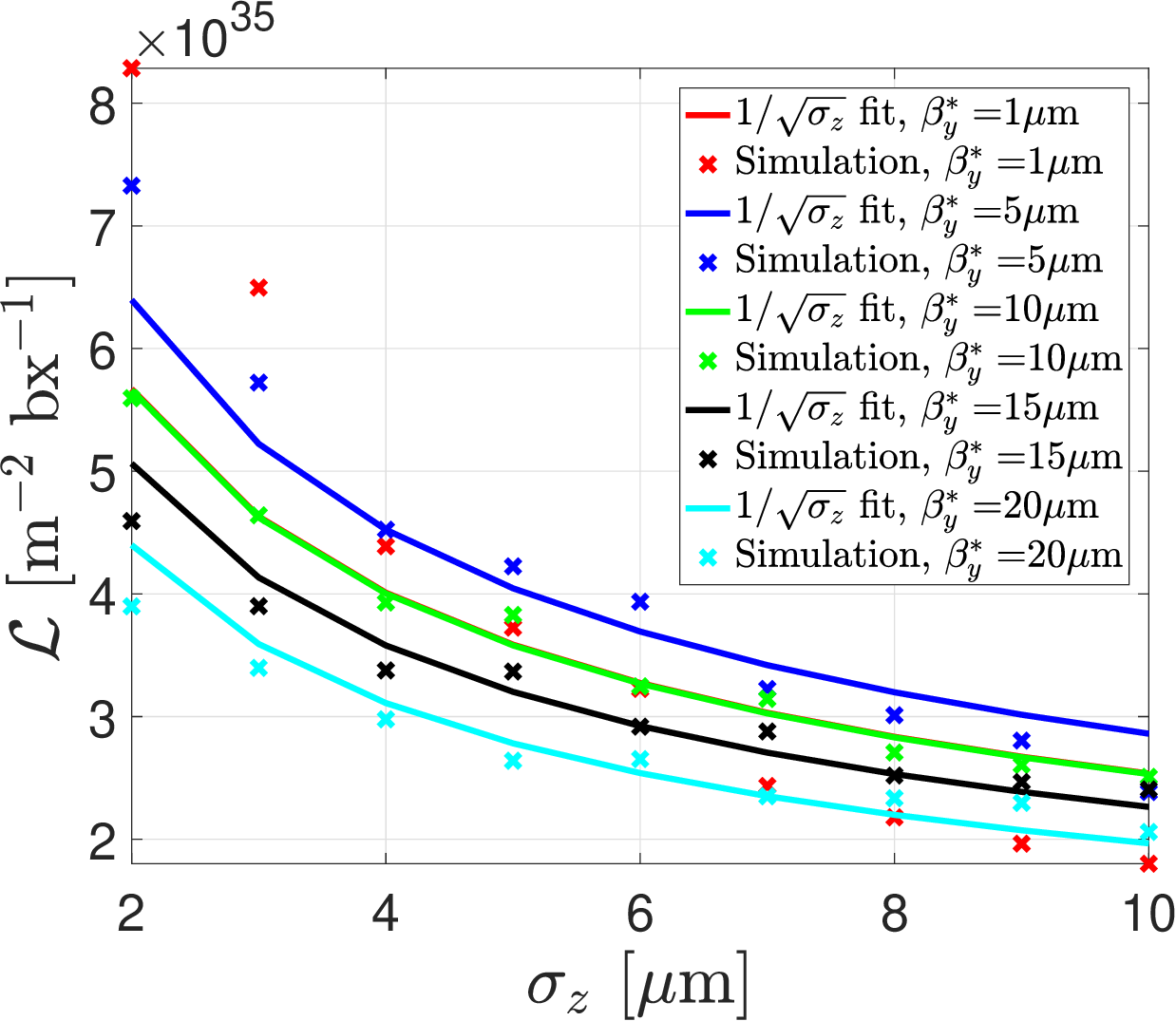}
    \caption{Total luminosity $\mathcal{L}$ vs. rms beam length $\sigma_z$ for several vertical beta functions $\beta_y$ along with corresponding theoretical $1/\sqrt{\sigma_z}$ fits.}
    \label{fig:2020_N5e9_sigmaZ_lumi_plot}
\end{figure}

Assuming $\sigma_x$ can be made sufficiently small despite technical constraints to keep $n_\gamma$ constant as $\sigma_z$ is reduced, and that $\sigma_y$ is kept constant, eq. \eqref{eq:luminosity_quantum_beamstrahlung} gives the scaling $\mathcal{L}\propto 1/\sqrt{\sigma_z}$. The luminosity is plotted against $\sigma_z$ for a selection of $\beta_y$ along with the corresponding $\mathcal{L}\propto 1/\sqrt{\sigma_z}$ fits in fig. \ref{fig:2020_N5e9_sigmaZ_lumi_plot}. 


The $1/\sqrt{\sigma_z}$-scaling agrees very well with simulation results, especially for larger values of $\beta_y$. The disagreement at small $\beta_y$ may be due to the hourglass effect, which imposes $\beta_y\geq\sigma_z$. When $\beta_y<\sigma_z$, a small beam size is only maintained over a small length, which reduces luminosity. Thus, using our range of $\sigma_z$-values, the luminosity appears to decrease faster than the $1/\sqrt{\sigma_z}$-scaling.


In a previous parameter study \cite{Chen2020modeling} on a $\SI{1.5}{\tera\electronvolt}$ PWFA linear accelerator, we found a parameter set for the main beam with acceptable stability, energy spread and efficiency. This parameter set involves an electron beam with $N=\SI{5e9}{}$ electrons and a rms beam length of $\sigma_z=\SI{5}{\micro\meter}$. The corresponding optimised results for a $N=\SI{5e9}{}$ $\sigma_z=\SI{5}{\micro\meter}$ beam from this study are listed in table \ref{tab:parameter} together with relevant results from \cite{Chen2020modeling}.



The normalised amplification factor $\Lambda/\Lambda_0$ \cite{normalizedAmplitude, Chen2020modeling} is a measure for stability used to quantify the amplification of the transverse jitter of the main beam. It is listed along with the relative rms energy spread $\sigma_\calE/\mean{\calE}$ and drive beam to main beam efficiency $\eta$ in table \ref{tab:parameter}. 

\begin{table}[!htb]
\centering
\caption{Main Parameters for a \SI{3}{\tera\electronvolt} PWFA Linear e\textsuperscript{$+$}e\textsuperscript{$-$} Collider}
    \begin{tabular}{lcc}
    \toprule
        \textbf{Parameter}            &   \textbf{Symbol [unit]}    &  \textbf{Value}      \\
    \midrule
    Plasma density  &   $n_0$ [\SI{e16}{\centi\meter^{-3}}] &   2.0\\
    Particle number     &$N$ [\SI{e9}{}]                        &5\\
    rms beam length     &$\sigma_z$ $[\SI{}{\micro\meter}]$                 &5\\
    
    Horizontal beta function  &\multirow{2}{*}{$\beta_x$ $[\SI{}{\milli\meter}]$}   &     \multirow{2}{*}{5}\\
    at IP\\ 
    
    Vertical beta function &\multirow{2}{*}{$\beta_y$ $[\SI{}{\micro\meter}]$}   &     \multirow{2}{*}{5}\\
    at IP\\ 
    
    Normalised horizontal  &\multirow{2}{*}{$\gamma\varepsilon_{x}$ $[\SI{}{\milli\meter \milli\radian}]$} &\multirow{2}{*}{0.887}\\
    emittance\\
    
    Normalised vertical &\multirow{2}{*}{$\gamma\varepsilon_{y}$ $[\SI{}{\milli\meter \milli\radian}]$} &\multirow{2}{*}{0.02}\\
    emittance\\
    
    
    Relative rms energy    &$\sigma_\calE/\mean{\calE}$ $[\%]$       &1.1\\
    spread\\
    
    Normalised amplification  &\multirow{2}{*}{$\Lambda/\Lambda_0$}                   &\multirow{2}{*}{6}\\
    factor\\
    
    Drive beam to main beam         &\multirow{2}{*}{$\eta$ $[\%]$}              &\multirow{2}{*}{37.5}\\
    efficiency\\ 
    
    Beam power/beam &   $P_\mathrm{b}/(f_\mathrm{r}n_\mathrm{b})$ [kWs] &   1.2\\
    Beamstrahlung photons/e\textsuperscript{$-$}  & $n_\gamma$  &   2.3\\
    
    Total luminosity/beam     &   $\mathcal{L}$    &   \multirow{2}{*}{4.3}\\
    crossing    &   [\SI{e35}{\meter^{-2}} bx\textsuperscript{$-1$}]   &\\
    
    Peak 1\% luminosity/beam    &   $\mathcal{L}_{0.01}$    &   \multirow{2}{*}{1.4}\\
    crossing  &   [\SI{e35}{\meter^{-2}} bx\textsuperscript{$-1$}]   &\\
    \bottomrule
    \end{tabular}
\label{tab:parameter}
\end{table}

In deriving this parameter set, we did not consider technological constraints on the vertical beta function. The vertical beta function $\beta_y=\SI{0.068}{\milli\meter}$ from the \SI{3}{\tera\electronvolt} CLIC parameter set \cite{CLIC_CDR} represents what is currently achievable, which is about one order of magnitude larger than our proposed value.

\subsection{Reduced Positron Beam Charge}

In the blow-out regime of PWFA, an electron beam will be focused by the positive ion background, while a positron beam will be defocused. Other acceleration schemes such as hollow channel plasma \cite{Gessner_hollowChannel} and the quasi-linear \cite{Doche_quasiLinear} regime have been proposed, but positron acceleration in plasma remains one of the main challenges in PWFA, and is considered to be significantly more challenging than electron acceleration.

Here we examine the effects of asymmetric e\textsuperscript{$+$}e\textsuperscript{$-$} collisions on luminosity at two energy levels, where the number of particles $N_\mathrm{e^+}$ in the e\textsuperscript{$+$} beam is only a fraction of the number of particles $N_\mathrm{e^-}$ in the e\textsuperscript{$-$} beam. Other beam parameters such as $\sigma_z$, $\beta_{x,y}$ and $\varepsilon_{\mathrm{N}x,y}$ are identical for the e\textsuperscript{$+$}e\textsuperscript{$-$} beams, and can be found in table \ref{tab:parameter}. The results for different scenarios are summarised in table \ref{tab:comparison_reduced_e+_3TeV} together with CLIC parameters \cite{CLIC_CDR}.


\begin{table*}[!hbt]
\centering
\caption{Parameter Comparison at \SI{3}{\tera\electronvolt} Collision Energy}
    \begin{tabular}{llcccc}
    \toprule
        \textbf{Parameter}  &   \textbf{Unit}  &   $N_{\mathrm{e}^+}=N_{\mathrm{e}^-}$  &   $N_{\mathrm{e}^+}=0.5N_{\mathrm{e}^-}$  &   $N_{\mathrm{e}^+}=0.1N_{\mathrm{e}^-}$  &   CLIC\\
    \midrule
    $N$ &   \SI{e9}{}   &   5/5 &   2.5/5   &   0.5/5   &  3.72/3.72\\
    
    $P_\mathrm{b}/(f_\mathrm{r}n_\mathrm{b})$   &   kWs   &   1.2/1.2   &    0.60/1.2   &    0.12/1.2    &   0.89/0.89\\
    
    $\calE_\mathrm{b}$  &   TeV &   1.5    &   1.5    &  1.5    &   1.5\\
    
    $\mathcal{L}$   &   \SI{e35}{\meter^{-2}} $\mathrm{bx}^{-1}$ &   4.26   & 1.89  & 0.32    &   0.38\\
    
    $\mathcal{L}_{0.01}$   &   \SI{e35}{\meter^{-2}} $\mathrm{bx}^{-1}$ &   1.41   & 0.80  & 0.17    &   0.13\\
    \bottomrule
    \end{tabular}
\label{tab:comparison_reduced_e+_3TeV}
\end{table*}

As a result of the reduced $N_\mathrm{e^+}$, the total luminosity is reduced by approximately the same factor compared to cases where $N_\mathrm{e^+}=N_\mathrm{e^-}$. However, by reducing $N_\mathrm{e^+}$, the beamstrahlung from the electron beam is also reduced, which results in a narrower luminosity spectrum. Alternatively, this also allows the horizontal beam size to be further reduced without increasing $n_\gamma$.

Even in the $N_\mathrm{e^+}=0.1N_\mathrm{e^-}$ scenario, a PWFA linear collider using the parameter set in table \ref{tab:parameter} can still provide a comparable luminosity level per beam crossing compared to CLIC\footnote{$n_\mathrm{b}=312$, $f_\mathrm{r}=\SI{50}{\hertz}$ for CLIC.}. Furthermore, the $N_\mathrm{e^+}=0.1N_\mathrm{e^-}$ scenario shown in table \ref{tab:comparison_reduced_e+_3TeV} has a luminosity spread that is significantly better than our defined tolerance of $\mathcal{L}/\mathcal{L}_{0.01}\approx 1/3$, which indicates that the horizontal beam size can likely be reduced even further to increase the total luminosity.

The muon collider submission to the European Particle Physics Strategy \cite{delahaye2019muon} stated that a \SI{14}{\tera\electronvolt} muon collider can provide a similar effective discovery potential as the \SI{100}{\tera\electronvolt} FCC. For comparison, the same parameters for \SI{14}{\tera\electronvolt} collision energy are shown in table \ref{tab:comparison_reduced_e+_14TeV}. 

\begin{table*}[!hbt]
\centering
\caption{Parameter Comparison at \SI{14}{\tera\electronvolt} Collision Energy}
    \begin{tabular}{llcccc}
    \toprule
        \textbf{Parameter}  &   \textbf{Unit}  &   $N_{\mathrm{e}^+}=N_{\mathrm{e}^-}$  &   $N_{\mathrm{e}^+}=0.5N_{\mathrm{e}^-}$  &   $N_{\mathrm{e}^+}=0.1N_{\mathrm{e}^-}$  &   CLIC\\
    \midrule
    $N$ &   \SI{e9}{}   &   5/5 &   2.5/5   &   0.5/5   &  3.72/3.72\\
    
    $P_\mathrm{b}/(f_\mathrm{r}n_\mathrm{b})$   &   kWs   &   5.61/5.61   &    2.80/5.61   &    0.56/5.61    &   4.17/4.17\\
    
    $\calE_\mathrm{b}$  &   TeV &   7.0    &   7.0    &  7.0    &   7.0\\
    
    $\mathcal{L}$   &   \SI{e35}{\meter^{-2}} $\mathrm{bx}^{-1}$ &   23.49   & 9.95  &    1.58 &   3.71\\
    
    $\mathcal{L}_{0.01}$   &   \SI{e35}{\meter^{-2}} $\mathrm{bx}^{-1}$ &   5.99   &  3.44    &    0.77 &   0.57\\
    \bottomrule
    \end{tabular}
\label{tab:comparison_reduced_e+_14TeV}
\end{table*}

At \SI{14}{\tera\electronvolt}, a common luminosity goal for linear colliders is \SI{40e34}{\per\centi\meter\squared\per\second}, which can be achieved in the $N_\mathrm{e^+}=N_\mathrm{e^-}$ and $N_\mathrm{e^+}=0.1N_\mathrm{e^-}$ scenarios with total beam powers of \SI{19}{\mega\watt} and \SI{156}{\mega\watt}, respectively. For comparison, the CLIC parameter set\footnote{Note that here the calculations in GUINEA-PIG were done with $\beta_{x,y}$ that do not take non-linear effects into account and thus differ from the values given in \cite{CLIC_CDR}. Here we chose $\beta_x=\SI{9.0}{\milli\meter}$ and $\beta_y=\SI{0.147}{\milli\meter}$, which are matched to the spot sizes and emittances given in \cite{CLIC_CDR} at \SI{3}{\tera\electronvolt}.} requires a total beam power of \SI{90}{\mega\watt} to achieve this luminosity. However, note that none of the parameter sets in table \ref{tab:comparison_reduced_e+_14TeV} have been optimised for \SI{14}{\tera\electronvolt}, which can be seen in the large luminosity spread. Furthermore, we also made the optimistic assumption that the same emittance levels and beta functions can be maintained for the two energy levels.

\section{CONCLUSION}




The suppression of beamstrahlung with decreasing beam length (given that the horizontal beam size can be scaled appropriately to limit beamstrahlung) is beneficial for PWFA, as short beams are preferred. The derived parameter set for a \SI{3}{\tera\electronvolt} e\textsuperscript{$+$}e\textsuperscript{$-$} PWFA linear collider shows a promising level of luminosity, while maintaining a reasonable luminosity spread. Even for a scenario where the positron beam only contains 10\% of the particles in the electron beam, the derived parameter set can still provide luminosity per beam crossing comparable to that of CLIC at \SI{3}{\tera\electronvolt}.

At \SI{14}{\tera\electronvolt}, the derived parameter set is able to achieve the luminosity goal of \SI{40e34}{\per\centi\meter\squared\per\second} with a significantly lower beam power than the CLIC parameter set. These parameters are however not optimised for this energy level, and thus give rise to a large luminosity spread.

The proposed parameter set furthermore has a vertical beta function that is an order of magnitude smaller than what is currently achievable. The CLIC parameter set has minimised the beam sizes by optimising the damping ring and the beam delivery system. Thus, further studies in reducing the beam sizes are required in order to achieve the luminosity goals.

Due to the challenges of accelerating positrons in a plasma, a $\gamma\gamma$ collider can be considered as an alternative to a e\textsuperscript{$+$}e\textsuperscript{$-$} collider. Furthermore, collisions of high-energy photons also have the potential to reach higher luminosities than e\textsuperscript{$+$}e\textsuperscript{$-$} collisions due to the absence of some beamstrahlung. A similar beam-beam study for a PWFA $\gamma\gamma$ collider will be conducted in future work.



%
%
\ifboolexpr{bool{jacowbiblatex}}%
	{\printbibliography}%

\begin{thebibliography}{99}   
	
	\bibitem{Blumenfeld}
	    I. Blumenfeld \textit{et al.},
	    `Energy doubling of 42 GeV electrons in a metre-scale plasma wakefield accelerator',
	    \textit{Nature},
	    vol. 445,
	    pp. 741--4,
	    Mar. 2007.
	
    \bibitem{Chen2020modeling}
        J. B. B. Chen, D. Schulte and E. Adli, 
        `Modeling and simulation of transverse wakefields in PWFA',
        \textit{J. Phys. Conf. Ser.},
        vol. 1596,
        p. 012057,
        Jul. 2020.

    \bibitem{Schulte_thesis}
        D. Schulte, 
        `Study of Electromagnetic and Hadronic Background in the Interaction Region of the TESLA Collider',
        PhD thesis,
        Hamburg U.,
        1997.
	
	\bibitem{accHandbook}
	    A. W. Chao, K. H. Mess, M. Tigner and F. Zimmermann.
        \textit{Handbook of Accelerator Physics and Engineering}.
        World Scientific Publishing Company,
        2013.
    
    \bibitem{BeamBeam_Yokoya}
        K. Yokoya and P. Chen.
        `Beam-beam phenomena in linear colliders',
        \textit{Lecture Notes in Physics},
        May 1995.
        
    \bibitem{normalizedAmplitude}
        A. Aksoy, D. Schulte and {\"O}. Yavas,
        `Beam dynamics simulation for the Compact Linear Collider drive-beam accelerator',
        \textit{Phys. Rev. Spec. Top. Accel. Beams},
        vol. 14, 
        084402. 8 p,
        Aug. 2011.

    \bibitem{CLIC_CDR}
        M. Aicheler \textit{et al.},
        \textit{A Multi-TeV Linear Collider Based on CLIC Technology: CLIC Conceptual Design Report},
        ser. CERN Yellow Reports: Monographs.
        Geneva: CERN,
        2012.

    \bibitem{Gessner_hollowChannel}
        S. Gessner \textit{et al.},
        `Demonstration of a positron beam-driven hollow channel plasma wakefield accelerator',
        \textit{Nature Communications},
        vol. 7,
        p. 11785,
        Jun 2016.

    \bibitem{Doche_quasiLinear}
        A. Doche \textit{et al.},
        `Acceleration of a trailing positron bunch in a plasma wakefield accelerator',
        \textit{Scientific Reports},
        vol. 7,
        Dec. 2017.
    
    \bibitem{delahaye2019muon}
        J. P. Delahaye \textit{et al.},
        \textit{Muon Colliders},
        2019.
        arXiV: \texttt{1901.06150}
    

	


	\end{thebibliography}
	{%

} 

%
%


\end{document}